\documentclass[a4paper,11pt]{article}
\begin{document}

\title{ Towards new background independent representations for 
Loop Quantum Gravity}

\author{Madhavan Varadarajan\\{\sl Raman Research Institute, Bangalore 560 080}\\
madhavan@rri.res.in}




\maketitle

\begin{abstract}
Recently, uniqueness theorems were constructed for the representation used in 
Loop Quantum Gravity. We explore the existence of alternate 
representations by weakening the assumptions of the so called LOST uniqueness
theorem. The weakened assumptions seem physically reasonable and retain the 
key requirement of  explicit background independence. For simplicity, we restrict attention
to the case of gauge group $U(1)$.
\end{abstract}

\section{Introduction}
Loop Quantum Gravity (LQG) 
is an attempt at canonical quantization of a
reformulation of classical gravity in terms of a spatial $SU(2)$ connnection
and its  conjugate electric field. The representation underlying LQG is one
in which the basic operators of the theory are holonomies of the connnection 
around spatial loops and electric fluxes through spatial surfaces. Much of the
progress in LQG and many of its results (such as the discrete spectra of 
spatial geometric operators) depend on the detailed properties of this 
representation. Hence it is important to know if the choice of this 
representation is essentially unique, given appropriate physically reasonable
requirements.
 
Work on this issue in recent years \cite{uniqueness} has culminated in the 
formulation and proof of beautiful uniqueness theorems by Fleischhack 
\cite{christian} and Lewandowski, Okolow, Sahlmann and Thiemann (LOST) 
\cite{lost}. Here we turn our attention to the LOST theorem.

The two key inputs for the LOST theorem
\footnote{Please see the reference \cite{lost} for a detailed description of
all the assumptions used in their proof.}
are as follows: \\
\noindent {\bf (1)} 
The algebra of holonomies and electric fluxes of 
which a representation is to be constructed. \\
\noindent {\bf (2)} The requirement that the representation contain a cyclic, 
spatially
diffeomorphism invariant state. 

While these are the simplest properties one may demand  from a putative
representation of a diffeomorphism invariant theory of $SU(2)$ connections and electric fields, one may 
enquire as to whether they are unduly restrictive. Let us discuss them one by 
one.

The holonomy- flux algebra of  {\bf (1)} agrees unambiguously 
with the corresponding 
Poisson bracket algebra except for the commutators of pairs of electric fluxes.
While one may expect these to vanish in classical theory, 
the analysis of Reference \cite{aaalexjose} shows that there are 
subtelities related to the lack of sufficient (functional) differentiability of the 
holonomy- flux variables which belie this
expectation. Motivated by this fact, LOST assume the commutators to take a 
particular form which is based on the interpretation of the flux operators
as derivations on a space of cylindrical functions \cite{lost}. While
these commutators, when evaluated in the representation used for LQG, 
have the physically appealing property of vanishing 
in a suitable ``large scale'', semiclassical approximation \cite{aaalexjose}, they are based on structures 
which do not appear in the classical phase space. Hence, the replacement of these 
commutators   
by any other reasonable choice which vanishes in a suitable 
``large scale'', semiclassical limit \cite{aaalexjose}, would not be unnatural.

Further, note that the algebra  involves only the holonomy- flux functions.
This seems sufficient since any other function of interest 
can be built as (limits of) sums and products of these functions, and 
presumably, the corresponding operators may be constructed by the 
corresponding (limits of) sums and products of the holonomy and flux operators.
However, in practice, in the representation currenty used in LQG, the
generators of spatial diffeomorphisms cannot be constructed in this manner
because the relevant limits do not exist in quantum theory. Indeed, only 
operators corresponding to finite spatial diffeomorphisms exist and these 
are {\em defined} through their natural action  on the flux- holonomy 
operators rather than constructed from the latter.
\footnote{The operators corresponding to finite spatial diffeomorphisms 
may also be constructed by their natural action on the space of
generalised connections. However, this recipe too,  does not derive from
a construction of limits of flux-holonomy operators.}
Hence it is natural to consider for the quantum theory, a {\em larger}
algebra 
generated by the holonomy, flux and 
finite spatial diffeomorphism operators. 
Clearly, we may only treat the 
finite spatial diffeomorphism operators as independent of the holonomy- flux 
ones  provided that, in the chosen Hibert space representation, these operators cannot be constructed as suitable limits 
of combinations of the holonomy - flux operators. We shall return to this 
point in the concluding section of this paper.

Let us call the enlarged algebra ${\cal UD}$ (in contrast to the algebra ${\cal U}$
defined by LOST in \cite{lost}). 
The following relations hold on the 
generators of the algebra  ${\cal UD}$ 
(these are, of course, in addition to the algebraic relations among
elements of the algebra ${\cal U}$):
\begin{eqnarray}
{\hat U}_{d_1}{\hat U}_{d_2}&=& {\hat U}_{d_1\circ d_2}, \label{ud1}\\
{\hat U}_{d}{\hat h}_{\alpha}{\hat U}^{\dagger}_{d}& =&
{\hat h}_{\alpha_d},\label{ud2} \\
{\hat U}_{d}{\hat E}_{{\cal S},f}{\hat U}^{\dagger}_{d}& =&
{\hat E}_{{\cal S}_d, f_d}, \label{ud3} \\
{\hat U}^{\dagger}_{d}&=& {\hat U}_{d^{-1}}.\label{ud4}
\end{eqnarray}
Here, ${\hat U}_{d},{\hat h}_{\alpha},{\hat E}_{{\cal S},f}$
refer to the operators corresponding to the finite diffeomorphism $d$,
the holonomy about a loop $\alpha$ and the flux through the surface 
${\cal S}$ smeared by the Lie algebra valued test field $f$ which has support on ${\cal S}$; 
${\alpha_d},{{\cal S}_d, f_d}$ refer to the 
images of ${\alpha},{{\cal S}, f}$ under the action of $d$. 
This completes our discussion of {\bf (1)}.

Next, consider the requirement {\bf (2)}. One consequence of {\bf (2)} is that 
the representation is cyclic. Cyclicity, by itself, is a weak assumption since
cyclicity is weaker than irreducibility and representations are usually
classified  in terms of their irreducible sectors. However, in light of the 
remarks in the previous paragraphs, 
it is not unnatural to consider cyclic representations of the 
enlarged algebra ${\cal UD}$ instead of the holonomy- flux algebra ${\cal U}$.

The further requirement  that the representation contain a 
diffeomorphism invariant state on ${\cal U}$ which is also cyclic ensures
that spatial diffeomorphisms act unitarily. It is 
quite natural to weaken this (rather strong) 
requirement to that of a unitary representation of
spatial diffeomorphisms irrespective of the existence of a diffeomorphism 
invariant state, whether cyclic or not.

In this work we explore the consequences of these weakened requirements 
in the context of $U(1)$ connections and electric fields.
We display a representation of ${\cal UD}$, 
inequivalent to the standard LQG type
representation appropriate to gauge group $U(1)$, in which spatial 
diffeomorphisms act unitarily. The restriction to $U(1)$ is for
simplicity.
There seems to be no reason as to why our general ideas should not go through for the 
$SU(2)$ case relevant to LQG.


The layout of the paper is as follows. A detailed description of
our constructions is presented in section 3.  Since  such an exposition
may obscure the structures essential to our construction, we
describe the broad underlying idea in section 2. This also serves to 
illustrate that there may exist implementations of our idea different from the
particular one chosen in section 3. Section 3 also  contains a discussion
of open issues. Section 4 contains our concluding remarks.
A key lemma is proved in the Appendix.

Depending on ones viewpoint on the validity of replacing 
${\cal U}$ by ${\cal UD}$, the particular representation displayed here
may be deemed to possess some undesirable features. We shall discuss this
in section 3.
However, as 
(at least to our knowledge) this is the first discussion of a representation
different from the standard one 
which supports the holonomy- flux operators as well as a 
unitary action of  spatial diffeomorphisms, at the very least our work
may be viewed
as initiating an exploration of background independent
representations inequivalent to the standard one. 

\section{A sketch of the underlying idea.}

Let $({\cal O}, {\cal H})$ denote a representation of the $*$- algebra 
${\cal O}$ on the non- seperable Hilbert space ${\cal H}$. The idea is to
construct a new representation of ${\cal O}$ in terms of 1 parameter families 
of states in ${\cal H}$. 
We  use the following notation.
$|\psi\rangle$ denotes an element of ${\cal H}$. 
A 1 parameter family of 
states in ${\cal H}$, 
$|\psi (a)\rangle \in {\cal H}\; \forall a \in R$
($R$ denotes the set of reals) is denoted by 
$\{|\psi (a)\rangle\}$.
The inner product between 
$|\psi_1\rangle,|\psi_2\rangle \in {\cal H}$
is 
$\langle \psi_1|\psi_2\rangle$. 
Given a 1 parameter family of states 
$\{|\psi (a)\rangle\}$, 
the state at parameter value $a=a_1$ is 
$|\psi (a_1)\rangle$. 
Since $|\psi (a_1)\rangle$ 
belongs to the 
1 parameter family of states 
$\{|\psi (a)\rangle\}$ 
we write 
 $|\psi (a_1)\rangle \in \{|\psi (a)\rangle\}$. 
Given a pair of 1 
parameter families of states 
$\{|\psi (a)\rangle\}$, $\{|\phi (b)\rangle\}$,
we denote the inner product between 
$|\psi (a_1)\rangle \in \{|\psi (a)\rangle\}$
and 
$|\phi (b_1)\rangle \in \{|\phi (b)\rangle\}$
by 
$\langle \psi (a_1)|\phi (b_1)\rangle$.

We are interested in a certain set of 1 parameter families of 
states in ${\cal H}$. Call this set ${\bf B}$. Elements of ${\bf B}$ 
consist of 1 parameter families of states subject to the 
following restrictions:\\

\noindent {\bf (i)}  Let 
$\{|\psi (a)\rangle\} \in {\bf B}$. We require that $\langle \psi (a_1)|\psi (a_2)\rangle =\delta_{a_1,a_2}$ where 
$\delta_{a_1,a_2}$ is the Kronecker delta function
\footnote{Roughly speaking, we aim to construct a new representation
in which the inner product is the Dirac delta function, $\delta (a_1, a_2)$
instead of the Kronecker delta function.
\label{dirac}
}
 which equals unity
when $a_1=a_2$ and vanishes otherwise.
Note that the existence of such 1 parameter families presupposes that 
${\cal H}$ is non-seperable. \\

\noindent {\bf (ii)}  Let 
$\{|\psi (a)\rangle\},\{|\phi (b)\rangle\} \in {\bf B}$. We require that
for every value of $a$ there exists at most one value of $b$ such that
$\langle \psi (a)|\phi (b)\rangle \neq 0$.\\

\noindent {\bf (iii)} Clearly {\bf (ii)} defines an invertible 
 function $b(a)$ from (a subset of) $R$ into $R$. We require that this
function be piecewise analytic i.e. the curve $(a, b(a))$ in the 
$a-b$ plane is required to be piecewise analytic.\\

\noindent {\bf (iv)} We also require that the ``overlap function''
$f(a): = \langle \psi (a)|\phi (b(a))\rangle$ is piecewise analytic.

We shall use elements of ${\bf B}$ to define a basis for the new 
representation. Specifically, every basis element is in correspondence 
with a pair of labels, $(A, \{|\psi (a)\rangle\})$. Here $A$ denotes a piecewise smooth scalar
of density $\frac{1}{2}$ on the real line and 
$\{|\psi (a)\rangle\} \in {\bf B}$.
\footnote{More precisely (as we shall see) it is the equivalence class of 
$\{|\psi (a)\rangle\}$ under reparametrizations rather than
$\{|\psi (a)\rangle\}$ itself which serves as an appropriate label.}
Next, we define the inner product between 
$(A, \{|\psi (a)\rangle\})$, $(B, \{|\phi (b)\rangle\})$ to be 
\begin{equation}
<A, \{|\psi (a)\rangle\}|B, \{|\phi (b)\rangle\}>=
\int_{\cal C} da 
\mid{\frac{db}{da}}\mid^{\frac{1}{2}}   A^*(a)
                      B(b(a)) \langle \psi (a)|\phi (b(a))\rangle
\label{ipheuristic1}
\end{equation}
Here ${\cal C}$ denotes the piecewise analytic curve $(a, b(a))$ (see {\bf (iii)}) above).
It is straightforward to verify that the inner product (\ref{ipheuristic1}) is hermitian
and that 
the density weights of $A,B$ imply that equation (\ref{ipheuristic1})
is independent of the specific parametrizations, $a$ and $b$,
of  $\{|\psi (a)\rangle\}$ and $\{|\phi (b)\rangle\}$. This suggests that
 we relabel our basis
elements by half densities and equivalence classes 
of 1 parameter families of states under reparameterization.  Accordingly,  denote the equivalence class of 1 parameter families
of states of which $\{|\psi (a)\rangle\}$
is a member, by $\Psi$, the set of basis states by ${\cal B}$ and elements of 
${\cal B}$  by $|A, \Psi\rangle$. Then the inner product between 
$|A, \Psi\rangle$ and $|B, \Phi\rangle$ is given by the right hand side of (\ref{ipheuristic1}) so that
\begin{equation}
\langle A, \Psi|B, \Phi\rangle =
\int_{\cal C} da 
\mid{\frac{db}{da}}\mid^{\frac{1}{2}}   A^*(a)
                      B(b(a)) \langle \psi (a)|\phi (b(a))\rangle
\label{ipheuristic2}
\end{equation}

Denote the new representation space by $V$. 
Any $|v\rangle \in V$ is a finite linear combination of elements of $\cal B$ so that
$|v\rangle = \sum_{I=1}^N c_I |A_I, \Psi_I\rangle$ where $c_I$ are complex coefficients.
We extend the scalar product (\ref{ipheuristic2}) to all of $V$ by appropriate (anti-)linearity.
Further, as suggested by equation (\ref{ipheuristic2}),  we make the following identifications in  $V$:
\begin{equation}
c|A, \Psi\rangle = |cA, \Psi\rangle
\label{identify1}
\end{equation}
 for all complex $c$ and $|A, \Psi\rangle \in {\cal B}$.
A final identification of vectors in $V$ is suggested by the linear structure of ${\cal H}$ and 
the scalar product (\ref{ipheuristic2}) as follows. 

Let 
$|A^{(i)}_{I_i}, \Psi^{(i)}_{I_i}\rangle,|B^{(j)}_{J_j}, \Phi^{(j)}_{J_j}\rangle \in {\cal B}, I_i=1,..,L_i,\;J_j=1,..M_j,\;i=1,2,\;j=1,2$ and let 
there exist the representatives
$\{|\psi^{(i)}_{I_i} (a)\rangle\}$,$\{|\phi^{(j)}_{J_j} (b)\rangle\}$ of 
$\Psi^{(i)}_{I_i}, \Phi^j_{J_j}$ in ${\bf B}$. Further, let 
$\sum_{I_1=1}^{L_1} A^{(1)}_{I_1}(a) |\psi^{(1)}_{I_1} (a)\rangle =\sum_{I_2=1}^{L_2} A^{(2)}_{I_2}(a) 
|\psi^{(2)}_{I_2} (a)\rangle$
and 
$\sum_{J_1=1}^{M_1} B^1_{J_1}(b) |\phi^{(1)}_{J_1} (b)\rangle =\sum_{J_2=1}^{M_2} B^{(2)}_{J_2}(b) 
|\phi^{(2)}_{J_2} (b)\rangle$.
Then, the identifications
\begin{equation}
\sum_{I_1=1}^{L_1} |A^{(1)}_{I_1}, \Psi^{(1)}_{I_1}\rangle =\sum_{I_2=1}^{L_2} |A^{(2)}_{I_2}, \Psi^{(2)}_{I_2}\rangle
\label{identify2a}
\end{equation}
and
\begin{equation}
\sum_{J_1=1}^{M_1} |B^{(1)}_{J_1}, \Phi^{(1)}_{J_1}\rangle =\sum_{J_2=1}^{M_2} |B^{(2)}_{J_2}, \Phi^{(2)}_{J_2}\rangle
\label{identify2b}
\end{equation}
are  suggested by 
the fact that the expression
\begin{equation}
c_{ij}:=\sum_{I_i=1}^{L_i} \langle A^{(i)}, \Psi^{(i)}_{I_i} | \sum_{J_j=1}^{N_j}B^{(j)}_{J_j}, \Psi^{(j)}_{J_j}\rangle 
\label{identity}
\end{equation}
is independent of $i,j$.
This fact follows straightforwardly from the scalar product (\ref{ipheuristic2}) and
the piecewise analyticity properties described in {\bf (iii)} and {\bf (iv)}.
Denote by $V_{new}$  the vector space obtained after  the identifications (\ref{identify1}), (\ref{identify2a}) 
have been made in $V$. $V_{new}$ serves as the representation space for our new representation.
Clearly, equation (\ref{ipheuristic2}) provides a hermitian  inner product on $V_{new}$. 
In the next section we shall show
that the particular choice of 1 parameter families of states in ${\cal H}$ used for abelian gauge theory ensures
that this inner product is also positive definite 
\footnote{Specifically, in section 3, the set ${\bf B}$ is such that any state
$|v\rangle = \sum_{I=1}^N c_I |A_I, \Psi_I\rangle$ can be rewritten using the equations 
(\ref{identify1}) and (\ref{identify2a}) as 
$|v\rangle = \sum_{I=1}^M d_J |B_J, \Phi_J\rangle$
with $\langle B_J, \Phi_J|B_K, \Phi_K\rangle = 0$ for $J\neq K$. 
Postive definiteness
then follows from equation (\ref{ipheuristic2}).
\label{fnoteposdef}
}
and may be used to complete $V_{new}$ to a Hilbert space
${\cal H}_{new}$.

Finally, we turn to the representation of the $*$- algebra ${\cal O}$ on $V_{new}$.
The action of any ${\hat O}\in {\cal O}$ on $|A,\{|\psi (a)\rangle\}\rangle$,
$ \{|\psi (a)\rangle\} \in {\bf B}$, is defined to be
$|A,\{{\hat O}|\psi (a)\rangle\}\rangle$. 
Here $\{{\hat O}|\psi (a)\rangle \in {\cal H} \;\forall a\}$  defines a new 1 parameter family of states.
While this family need not satisfy {\bf (i)}- {\bf (iv)}, we assume that it can be decomposed into a finite linear
combination of 1 parameter families which do. Thus we assume that 
\begin{equation}
{\hat O}|\psi (a)\rangle=  \sum_{I=1}^N A_I(a)|\psi_I (a)\rangle,
\label{assumeo}
\end{equation}
where $\{|\psi_I (a)\rangle\} \in {\bf B}$ and $A_I$ are piecewise smooth scalar functions (of weight zero).
We set $|A,\{{\hat O}|\psi (a)\rangle\}\rangle = \sum_{I=1}^N|AA_I,\{|\psi_I (a)\rangle\}\rangle$
so that 
\begin{equation}
{\hat O}|A, \Psi\rangle := \sum_{I=1}^N |AA_I, \Psi_I\rangle
\label{defoheuristic}
\end{equation}
 in obvious notation. 
It follows from equation (\ref{identify2a}) that the definition (\ref{defoheuristic}) 
is independent  of the particular decomposition of 
${\hat O}|A, \Psi\rangle$ used and that it, in conjunction with the inner product 
(\ref{ipheuristic2}) provides a $*$- representation of the $*$- algebra ${\cal O}$ on $V_{new}$.

In the next section, we shall provide a precise implementation of these ideas for abelian 
gauge theory. Before we conclude this section, we mention a useful heuristic which serves to 
emphasize the point made in Footnote \ref{dirac}. This is not required for the rest of the paper
and the reader may proceed straight to section 2 if it so desires.
Define $|\psi_{new} (a)\rangle$,
$|\phi_{new} (b)\rangle$   through 
\begin{eqnarray}
|A,\Psi\rangle &=:& \int da A(a)|\psi_{new} (a)\rangle \\
|B,\Phi\rangle &=:& \int db B(b)|\phi_{new} (b)\rangle 
\end{eqnarray}
with the new inner product being 
\begin{equation}
\langle\psi_{new} (a)|\phi_{new} (b)\rangle>= 
\langle\psi (a)|\phi (b))\rangle \delta (b, b(a)) \mid \frac{db}{da}\mid^{\frac{1}{2}}
\label{diracip}
\end{equation}
In this notation we have
\begin{equation}
\langle A, \Psi| =: \int da A^*(a)\langle \psi_{new} (a)|
\end{equation}
so that 
\begin{equation}
\langle A,\Psi|B, \Phi\rangle:= \int da db A^*(a) B(b) \langle\psi_{new} (a)|\phi_{new} (b)\rangle>
\end{equation}
which reduces to equation (\ref{ipheuristic2}) if we use the definition (\ref{diracip}). This notation brings 
out the Dirac delta normalization mentioned in Footnote \ref{dirac} and easily lends itself to the 
linearity based identifications (\ref{identify1}) and (\ref{identify2a}).

\section{The new representation for the case of gauge group $U(1)$.}

We set ${\cal O}$ to be the $*$- algebra ${\cal UD}$ (see the discussion associated with the equations 
(\ref{ud1})- (\ref{ud4})) and $({\cal O},{\cal H})$ to be the standard ``flux network'' representation
\cite{meqef,aajurekflux} which is the abelian analog of the spin network representation
currently used in LQG.

We provide a brief review of the flux network representation in section 3.1. In section 3.2, we define the 
set ${\bf B}$ of 1 parameter families of states on which the new representation is based and show that
 the inner product
(\ref{ipheuristic2}) is positive definite on the representation space $V_{new}$. In section 3.3, we display the action 
of the basic operators ${\hat h}_{\alpha}, {\hat E}(f), {\hat U}_d$ and show that the assumption (\ref{assumeo})
is valid. As mentioned in section 2, this ensures that ${\cal UD}$ is represented on ${\cal H}_{new}$.

\subsection{Review of the $U(1)$ flux network representation.}

This section provides a brief review of the flux network representation for a
difffeomorphism invariant  theory of
$U(1)$ connections and conjugate electric fields. Our primary aim is to establish notation.
We refer the reader to the review
article \cite{aajurekreview} and the references contained therein for a complete presentation.

Let $\Sigma$ be a 3 dimensional, compact, real analytic manifold without boundary. 
The phase space variables are a $U(1)$ connection $A_a(x)$ and its conjugate 
(unit density weight) electric field $E^a(x)$. 
Let $\alpha$ be a closed oriented graph composed of closed analytic edges $e_J, J=1,..,M$.
Let each edge $e_J$ be labelled by an integer $p_J$ such that at each vertex the sum of integers 
labelling outgoing edges equals that for incoming edges. The holonomy of the connection associated
with this labelled graph is $h_{\alpha, {\vec p}} = \exp (i\sum_{J=1}^M \int_{e_J}A_adx^a)$. As shown in 
Reference \cite{meqef} these graph holonomies are in correspondence with the more commonly used
loop holonomies i.e. for every such labelled graph $\alpha, {\vec p}$ there exists a loop $\beta$ such
that $h_{\alpha, {\vec p}}= h_{\beta}$ where $h_{\beta}:= \exp (i\oint_{\beta} A_adx^a)$ is the usual
loop holonomy.

The 
smeared electric flux through a 2 dimensional   surface ${\cal S}$ is $E_{{\cal S},f}= \int_{\cal S}f(x) E^a(x)d^2s_a$.
Here $f(x)$ is a smooth function of compact support on ${\cal S}$. The surface ${\cal S}$ is
chosen as in Reference \cite{aaalexjose}. For our purposes it is important to note that 
this choice is such that the surface is semianalytic (Please see Reference \cite{lost} for a defintion of
semianalyticity.). 
The holonomy- flux Poisson bracket algebra leads
unambiguously to the following commutators in quantum theory:
\begin{equation}
[{\hat h}_{\alpha_1, {\vec p}_1}, {\hat h}_{\alpha_2, {\vec p}_2}]=0, \;\;\;\;\;\;\; 
[{\hat h}_{\alpha, {\vec p}}, {\hat E}_{{\cal S},f}] = -\hbar \big( \sum_i f(x_i) \sum_{I_i}p_{I_i}\kappa_{I_i}\big)
{\hat h}_{\alpha, {\vec p}},
\label{pb1}
\end{equation}
where $i$ labels the transverse 
intersections of $\alpha$ with ${\cal S}$
(by which we mean that at least one edge of $\alpha$ is transverse to ${\cal S}$ at the intersection point), 
$I_i$ ranges over the edges of
$\alpha$ which intersect ${\cal S}$ transeversely at the point $x_i$ and $\kappa_{I_i}= 1$ or $-1$ depending on the
relative orientations and positions of $e_{I_i}$ and ${\cal S}$ (it turns out that without loss of generality
one can always arrange for $x_i$ to be a vertex of $\alpha$; see Footnote \ref{fine} and References 
\cite{aajurekarea,meqef}). Finally, the commutator of a pair of fluxes is
\begin{equation} 
[{\hat E}_{{\cal S}_1,f_1},{\hat E}_{{\cal S}_2,f_2}]= 0 .
\label{pb2}
\end{equation}
We emphasize here that in the case of $U(1)$ gauge group, the commutator of 
two electric flux variables (\ref{pb2}) vanishes unambiguously \cite{aaalexjose} and the 
subtlety mentioned in the Introduction does not arise. The holonomy and flux operators
together with their commutators (\ref{pb1})- (\ref{pb2}) generate the algebra $\cal{U}$  on 
which the $*$- relations are induced from the adjointness properties of the holonomy- flux operators:
\begin{equation}
({\hat h}_{\alpha , {\vec p}})^{\dagger}=  ({\hat h}_{\alpha , {\vec p}})^{-1} \;\;\;\;\;\;\;
({\hat E}_{{\cal S},f})^{\dagger}= {\hat E}_{{\cal S},f}.
\label{adjointness}
\end{equation}
The $*$- algebra ${\cal U}$ is represented on the Hilbert space ${\cal H}$ which is spanned by an 
orthonormal basis of flux network states. 
For simplicity, we shall restrict attention to gauge invariant states.
Each such state (with the exception of the state $|\circ\rangle$ defined below) 
is in correspondence with a closed oriented 
graph, every edge of which is labelled by a (non- trivial) representation of the Lie algebra of $U(1)$ i.e. by a 
non- zero integer. Each edge is required to be analytic and closed and at each vertex the sum of the integer labels of 
the incoming edges equals that of the outgoing edges.  Consider such a graph $\gamma$ with 
edges $e_I,\; I=1,..,N$, each labelled by a non-zero  integer $n_I$. The corresponding flux network state is 
denoted by $|\gamma, {\vec n}\rangle$. The flux network state associated with the trivial graph (with no edges) is
denoted by $|\circ\rangle$.

The action of the holonomy operator, ${\hat h}_{\alpha , {\vec p}}$,  on the 
flux network state $|\gamma, {\vec n}\rangle$ is 
\begin{equation}
{\hat h}_{\alpha , {\vec p}}|\gamma, {\vec n}\rangle
= |\gamma \cup  \alpha , {\vec n}\cup{\vec p}  \rangle .
\label{holonomyaction}
\end{equation}
The labels on the right hand side are defined as follows.
Consider any (closed, oriented) graph (with closed analytic edges) finer than $\gamma$ and $\alpha$.
\footnote{A graph $\gamma_1$ is said to be finer than a graph $\gamma_2$ iff every edge of the latter can be composed
of edges in the former. We may also refer to $\gamma_2$ as being coarser than $\gamma_1$. A flux network state, strictly
speaking, is labelled by an equivalence class of graphs and integer labellings where the pair $\gamma_1, {\vec n}_1$
and the  pair $\gamma_2, {\vec n}_2$ are equivalent if 
the images of $\gamma_1$ and $\gamma_2$ in $\Sigma$ are identical and either
(a)$\gamma_1$ is finer than $\gamma_2$ and the edges in the former
which compose to yield an edge in the latter are all  labelled by the (same) integer which labels the edge in the latter
or (b) vice versa . We also note that flipping the orientation of edges in a flux network state is the same as retaining 
the orientation and flipping the sign of the integer labels.
\label{fine}
}
Any edge $e$ of this graph is labelled as follows (below $e_I$ denotes the $I$th edge of $\gamma$ 
and $e^{\alpha}_J$ the Jth edge of $\alpha$ ):\\
\noindent (a) If $e\subset e_I, e\subset e^{\alpha}_J$ and $e,e_I,e^{\alpha}_J$ have the same orientation,
$e$ is assigned the label $n_I+p_J$. In case of orientation mismatches with 
only $e_I$, only $e^{\alpha}_J$ or both $e_I$ and $e^{\alpha}_J$ the label is $-n_I+p_J$, $n_I-p_J$ or $-n_I-p_J$. \\
\noindent (b) If  $e\subset e_I$ and $e$ intersects $\alpha$ at most at isolated points, 
$e$ is assigned the label $n_I$ if its orientation is the same as that induced from $e_I$ else it is labelled $-n_I$. \\
\noindent (c) If $e\subset e^{\alpha}_J$ and $e$ intersects $\gamma$ at most at isolated points,
then $e$ is assigned the label $p_J$ if its 
orientation is the same as that of $e^{\alpha}_J$ (else it is labelled $-p_J$).\\
\noindent The pair $\gamma \cup  \alpha , {\vec n}\cup {\vec p}$ 
is defined by (a)- (c) above with the additional caveat that any edge
$e$ whose integer label vanishes as a result of (a) is dropped from the graph.

The action of the electric flux operator ${\hat E}_{{\cal S}, f}$ on the state $|\gamma, {\vec n}\rangle$ is 
\begin{equation}
{\hat E}_{{\cal S}, f}
|\gamma, {\vec n}\rangle
=  \hbar \big( 
\sum_i f(x_i) \sum_{I_i}n_{I_i}\kappa_{I_i} \big)
|\gamma , {\vec n}\rangle ,
\label{fluxaction}
\end{equation}
where $x_i, I_i, \kappa_{I_i}$ are defined  as in  equation (\ref{pb1}).
The operator ${\hat U}_d$ which corresponds to the action of the finite diffeomorphism
$d$ (see equations (\ref{ud1})- (\ref{ud4})) acts on  the state $|\gamma, {\vec n}\rangle$ as 
\begin{equation}
{\hat U}_d
|\gamma, {\vec n}\rangle
=  |\gamma_d , {\vec n}\rangle ,
\label{diffaction}
\end{equation}
where on the right hand side, $\gamma_d$ is the image of the graph $\gamma$ under the diffeomorphism $d$ and 
the image of the edge $e_I$ of the graph $\gamma$ under $d$ is labelled by $n_I$.
We shall restrict attention to analytic diffeomorphisms $d$, so that $\gamma_d$ is also a piecwise analytic graph.

In addition to their action on $|\gamma, {\vec n}\rangle$  (\ref{holonomyaction})- (\ref{diffaction}), the above operators act as follows 
on the trivial graph state.
\begin{equation}
{\hat h}_{\alpha , {\vec p}}|\circ\rangle
= |\alpha , {\vec p}\rangle ,
\label{holonomyactiontrivial}
\end{equation}
We also have 
\begin{equation}
{\hat E}_{{\cal S},f}
|\circ\rangle
=  0, 
\label{fluxactiontrivial}
\end{equation}
and
\begin{equation}
{\hat U}_d
|\circ\rangle
=  |\circ\rangle ,
\label{diffactiontrivial}
\end{equation}
 
It can be verified that the equations (\ref{holonomyaction})- (\ref{diffactiontrivial}) together with the orthonormality of
the flux network basis, provide a $*$- representation for the $*$ - algebra ${\cal UD}$ (\ref{ud1})- (\ref{ud4}) with 
$\cal U$ defined through (\ref{pb1})- (\ref{pb2}) and (\ref{adjointness}).

\subsection{Construction of the  new Hilbert space.}

We define the set ${\bf B}$ of 1 parameter families of states as follows. Let $\gamma$ be a closed graph with 
closed analytic edges $e_I, I=1,..,N$. Let $\xi$ be a real analytic vector field on $\Sigma$ and $U$ an open 
neighbourhood in $\Sigma$. Let $\phi_{\xi} (s),\; s\in R$, denote the one 
parameter family of diffeomorphisms of $\Sigma$ generated by $\xi$ with 
$\phi_{\xi} (0)$ being the identity map. Let $\gamma (\xi ,s)$ be the graph obtained by the action of 
$\phi_{\xi} (s)$ on $\gamma$ i.e. $\gamma (\xi ,s) =\phi_{\xi} (s)\gamma$. We subject $\gamma, \xi, U$ to the 
following restrictions:\\
\noindent (a) We require that each $e_I$ admits a non-self intersecting, open, analytic extension 
${\tilde e}_I$ such that ${\tilde e}_I \subset U$. \\
\noindent (b) We require that $\xi$ be nonvanishing in $U$ and transverse to every ${\tilde e}_I, I=1,..,N$.

Clearly, there exists an open neighbourhood $S$ of the origin such that for every $s,s^{\prime},s^{\prime\prime} \in S$,
\begin{equation}
\phi_{\xi} (s){\tilde e}_I \subset U, \;\;\;\;\;I=1,..,N,
\label{3.2.1}
\end{equation}
\begin{equation}
\gamma (\xi , s^{\prime}) = \gamma (\xi, s^{\prime\prime})\;\;\; {\rm iff}\;\; s^{\prime}= s^{\prime\prime}.
\label{3.2.2}
\end{equation}

Consider a flux network $|\gamma, {\vec n}\rangle$ based on the graph $\gamma$. Let 
$|\gamma(\xi , s), {\vec n}\rangle$ denote the flux network based on $\gamma (\xi ,s)$ such that the $I$th edge
of $\gamma (\xi ,s)$, namely $\phi_{\xi} (s)e_I$, is labelled by $n_I$. 
Finally, let ${\bf S}\subset S$ be a closed interval
containing the origin. Then the set ${\bf B}$ consists of the 1 parameter families of states
$\{\hat{h}_{\alpha ,{\vec p}}|\gamma(\xi ,s), {\vec n}\rangle,\; s\in {\bf S}\}$ for all possible choices of 
$\alpha, {\vec p}, \gamma, {\vec n}, \xi , {\bf S}$. If $\alpha$ is chosen to be the trivial graph, $\circ$, there is
no labelling ${\vec p}$ and
the 1 parameter family of states is just $\{|\gamma(\xi ,s), {\vec n}\rangle,\; s\in {\bf S}\}$.

Clearly, equation (\ref{3.2.2}), in conjunction with the orthogonality of flux network states based on different graphs,
ensures that {\bf (i)},{\bf (ii)} of section 2 hold. We now show that {\bf (iii)} also holds.
Let $\{\hat{h}_{\alpha_i ,{\vec p}_i}|\gamma_i(\xi_i ,s_i), {\vec n_i}\rangle,\; s_i\in {\bf S}_i\},\;i=1,2$ be a pair of states 
such that infinitely many points $(s_1, s_2(s_1))$ exist where the states are non- orthogonal (here $\alpha_i$ could
also be the trivial graph in which case the labelling ${\vec p}_i$ is absent).
Specifically, let $s_{1min}$and $s_{1max}$ be the minimum and maximum values of $s_1$ for which $s_2(s_1)$ exists.
Then there are infinitely many points $s_1, s_1 \in [s_{1min},s_{1max}]$ such that $s_2(s_1)$ exists.
Next, note that
the images of  
of $\gamma_1(\xi_1,s_1),\gamma_2(\xi_2,s_2(s_1))$ agree in $\Sigma$. Hence,  we choose the two graphs to have the
same number of edges (That this does not entail any loss of generality follows from Footnote \ref{fine}.).
Let the $I$th edge of $\gamma (\xi_1, s_1)$ (i.e. $\phi_{\xi_1} (s_1)e_{1I}$) agree with the $J$th edge of
$\gamma (\xi_2, s_2(s_1))$ (i.e. $\phi_{\xi_2} (s_2(s_1))e_{2J}$). While in principle $J$ could be a function 
of both $I$ and $s_1$, the fact that $\gamma_1, \gamma_2$ have a finite number of edges ensures that for
infinitely many points $(s_1, s_2(s_1))$ there exists $J= J(I), I=1,..,N$  independent of $s_1$.
An application of Lemma 1 of the Appendix to the edges ${\tilde e}_{1I},{\tilde e}_{2J(I)}$ shows that  the surfaces 
${\cal S}_{1I}, {\cal S}_{2J}$ generated by the
action of $\phi_{\xi_1} (s_1),\phi_{\xi_2} (s_2)$ on the edges $e_{1I}, e_{2J(I)}$ are analytic with 
analytic charts $(t_{1I},s_1),(t_{2J(I)},s_2)$ where $t_{1I}, t_{2J(I)}$ are analytic parameterizations of 
${\tilde e}_{1I}, {\tilde e}_{2J(I)}$. Since these surfaces intersect at infinitely many curves, the portion of
${\cal S}_{1I}$ 
between the curves $s_1= s_{1min}$ and $s_1= s_{1max}$
must coincide with the portion of ${\cal S}_{2J(I)}$ 
between the curves $s_2 (s_{1min})$ and 
$s_2 (s_{1max})$. Then it follows from the analyticity of the charts mentioned above that
$s_2(s_1)$ is an analytic function and hence that {\bf (iii)} of section 2 holds.
\footnote{It is not difficult to see (for example via a straightforward application of the Cauchy- Kowalewski theorem 
\cite{CKthm}
to a 3 dimensional chart obtained by dragging a 2 dimensional patch along the orbits of a suitably defined analytic
vector field) that the fact that $(t_{1I}, s_1)$ and $(t_{2J(I)}, s_2)$ define analytic charts on a common analytic
surface implies that the latter are analytic functions of the former. In particular, $s_2$ is an analytic function
of $t_{1I}$ and $s_1$; however the fact that the analytic function 
$s_2$ is independent of $t_{1I}$ for infinitely many values of $s_1$
implies that $s_2$ is an analytic function only of $s_1$.}

Finally, it follows from the orthonormality of the flux network states that the relevant ``overlap function''
of {\bf (iv)} is piecewise constant and hence piecewise analytic.

The next step is to define the set ${\cal B}$. In what follows, we will drop the label $\xi$ from the
set of labels characterising states in ${\bf B}$ and it will be understood that any  1 parameter family
of graphs, $\gamma (s)$, has been obtained by dragging the graph $\gamma$ along the orbits of 
some analytic vector field in the manner discussed above. As in section 2, basis elements for our new representation
are in correspondence with an element of ${\bf B}$ together with a piecewise smooth 
scalar density of weight $\frac{1}{2}$.
We designate such elements by $|C,\alpha,{\vec p}, \{\gamma (s),{\vec n}, s\in {\bf S}\}\rangle$ 
(and by $|C,\{\gamma (s),{\vec n}, s\in {\bf S}\}\rangle$ for the case $\alpha = \circ$), 
where $C$ is a scalar half density
of compact support such that $C(s)$ is supported in ${\bf S}$. 
As in section 2, two such elements are to be identified if they  are related by analytic reparametrisations.
Specifically, the  set $(C, \gamma(s), s\in {\bf S})$ is equivalent to the set 
$(C, \gamma^{\prime} (s^{\prime}), s^{\prime} 
\in {\bf S}^{\prime})$
if $s(s^{\prime})$ is an analytic bijection from ${\bf S}^{\prime}$ to  ${\bf S}$ such that
\begin{equation}
\gamma^{\prime} (s^{\prime}) = \gamma (s(s^{\prime})), \;\;\;\forall s^{\prime} \in {\bf S}^{\prime}.
\end{equation} 
In addition, since $C$ is a half density, we have the
\begin{equation}
C(s^{\prime}) = C(s(s^{\prime})) |ds/ds^{\prime}|^{\frac{1}{2}}
\end{equation}
with $C(s^{\prime})$ being supported in ${\bf S}^{\prime}$.
We refer to the equivalence class of which $|C, \alpha,{\vec p},\{\gamma (s),{\vec n}, s\in {\bf S}\}\rangle$ 
is a member
\footnote{Strictly speaking this equivalence class should contain only elements of ${\bf B}$ which 
means that only those analytic reparameterizations should be permitted for which the parameter
takes values in a closed set {\em containing the origin}. However, we find it convenient to 
admit all possible analytic reparametrisations. Hence given an element of ${\bf B}$, its equivalence class
contains all 1 parameter families which are related to it via arbitrary analytic reparameterizations.
Thus, if ${\bf S}$ in the discussion above contains the origin, it is not necessary that ${\bf S}^{\prime}$
does.}
by $|C, \Gamma, \alpha,{\vec p},{\vec n}\rangle$
 and that of 
 $|C, \{\gamma (s),{\vec n}, s\in {\bf S}\}\rangle$ by $|C, \Gamma, {\vec n}\rangle$.
 Here, the information in ${\bf S}$ is implicit in the 
support of the half density $C$. Thus, the set ${\cal B}$ is composed of elements of the 
form $|C,\alpha,{\vec p}, \Gamma, {\vec n}\rangle$, $|C, \Gamma, {\vec n}\rangle$.
The inner product (\ref{ipheuristic2})  between
$|C_1,\alpha_1,{\vec p}_1, \Gamma_1, {\vec n}_1\rangle,$ 
and $|C_2,\alpha_2,{\vec p}_2, \Gamma_2, {\vec n}_2\rangle,$ is  
\begin{eqnarray}
\langle C_1, \alpha_1,{\vec p}_1,\Gamma_1,{\vec n}_1|C_2,\alpha_2,{\vec p}_2, \Gamma_2, {\vec n}_2\rangle =
\;\;\;\;\;\;\;\;\;\;\;\;\;\;\;\;\;\;\;\;\;\;\;\;\;\;&\nonumber\\
\int_{\cal C} ds_1 
\mid{\frac{ds_2}{ds_1}}\mid^{\frac{1}{2}}   C_1^*(s_1)
                      C_2(s_2(s_1)) 
\langle \gamma_1 (s_1)\cup \alpha_1, {\vec n}_1\cup {\vec p}_1|\gamma (s_2(s_1))\cup \alpha_2, 
{\vec n}_2\cup {\vec p}_2 
\rangle & .
\label{ip1}
\end{eqnarray}
Clearly, on the curve ${\cal C}$ it must be the case that 
the flux networks $|\gamma_1 (s_1)\cup \alpha_1 , {\vec n}_1\cup {\vec p}_1\rangle$ 
and 
$|\gamma (s_2(s_1))\cup \alpha_2, {\vec n}_2\cup {\vec p}_2\rangle$ 
are identical so that the equation (\ref{ip1}) reduces to
\begin{equation}
\langle C_1, \alpha_1, {\vec p}_1, \Gamma_1,{\vec n}_1|C_2,\alpha_2, {\vec p}_2, \Gamma_2, {\vec n}_2\rangle =
\int_{\cal C} ds_1 
\mid{\frac{ds_2}{ds_1}}\mid^{\frac{1}{2}}   C_1^*(s_1)
                      C_2(s_2(s_1)) .
\label{ip2}
\end{equation}
It is also straightforward to verify that the same equation holds with the
left hand side replaced by $\langle C_1, \Gamma_1,{\vec n}_1|C_2, \Gamma_2, {\vec n}_2\rangle$.
The next step is use the elements of ${\cal B}$ to generate the vector space $V_{new}$ subject to the 
identifications (\ref{identify1}) and (\ref{identify2a}). The inner product (\ref{ip1}) is extended
to  $V_{new}$ by appropriate linearity and antilinearity.
The general considerations of section 2 show that this inner product is Hermitian on $V_{new}$. We now 
demonstrate that it is also positive definite.

We present the argument for positive definiteness in a form readily generalizable to the non- abelian case.
Our strategy (along the lines of Footnote \ref{fnoteposdef}) is to rewrite  any  linear combination of
elements belonging to ${\cal B}$ as  one in which each term is based on a 1 parameter family of graphs
such that for any pair of terms, the corresponding pair of 1 parameter families of graphs have no common
images in $\Sigma$. Let $|v\rangle = \sum_{i=1}^m |\Psi_i\rangle \in V_{new}$ where
\begin{equation}
|\Psi_i\rangle = | C_i,\alpha_i, {\vec p}_i , \Gamma_i ,{\vec n}_i \rangle \in {\cal B} 
\end{equation}
Let $j$ be such that $\langle \Psi_1|\Psi_j\rangle \neq 0$ so that $\gamma_j(s_j)$ agrees with $\gamma_1(s_1)$
(almost) everywhere
along the curve ${\cal C}_{1j}$ in the $s_1-\;s_j$ plane. Using  equation (\ref{identify2a}) and an appropriate
analytic reparametrization, we have that 
\begin{equation}
| C_j, \alpha_j , {\vec p}_j, \Gamma_j ,{\vec n}_j \rangle := |C_{1j},\alpha_j , {\vec p}_j, \Gamma_1, {\vec n}_j\rangle 
                      + |C^{\prime}_j,\alpha_j , {\vec p}_j, \Gamma_j,{\vec n}_j\rangle ,
\end{equation}
where
\begin{eqnarray}
C_{1j}(s_1) &=& C_j(s_j(s_1)) \mid{\frac{ds_j}{ds_1}}\mid^{\frac{1}{2}}
\;\; \forall s_1 \;\;{\rm such}\;{\rm that}\; s_j(s_1)\;{\rm exists}, \\
&=&  0 \;\;\;{\rm elsewhere},\\
C^{\prime}_j(s_j) &=& 0 \;\; \forall s_j\;\;{\rm such}\;{\rm that}\; (s_1,s_j)\in {\cal C}_{1j} \\
&=& C_j(s_j) \;\;\;{\rm elsewhere}.
\end{eqnarray}
Next, define, for $i\neq 1$,
\begin{eqnarray}
|\Psi^{\prime}_i\rangle &=& |C^{\prime}_i,\alpha_i , {\vec p}_i, 
\Gamma_i, {\vec n}_i\rangle \;\;{\rm if}\;{\cal C}_{1i}\;{\rm exists},\\
&=& |C_i,\alpha_i , {\vec p}_i,\Gamma_i,{\vec n}_i\rangle \;\;{\rm otherwise}\\
\end{eqnarray}
and
\begin{equation}
|\Psi^{\prime}_1> = |C_1+ \sum_{j}C_{1j},\alpha_1 , {\vec p}_1, \Gamma_1, {\vec n}_1\rangle,
\end{equation}
where the sum is over all $j$ such that the curve ${\cal C}_{1j}$ exists. Then using equation (\ref{identify2a}),
we have that $|v\rangle = \sum_{i=1}^m |\Psi^{\prime}_i\rangle$ where $\langle \Psi^{\prime}_1|\Psi^{\prime}_i\rangle= 0$
for all $i\neq 1$. Next, apply this procedure to the sum $\sum_{i=2}^m |\Psi^{\prime}_i\rangle$ so as to ``orthogonalise''
with respect to $|\Psi^{\prime}_2\rangle $. 
Clearly, by repeating this procedure enough times, we may rewrite $|v\rangle$
in the form
\begin{equation}
|v\rangle = \sum_{i=1}^m |D_i,\alpha_i , {\vec p}_i, \Gamma_i, {\vec n}_i\rangle
\label{ognlsum}
\end{equation}
such that $\langle D_i,\alpha_i , {\vec p}_i, \Gamma_i, {\vec n}_i|D_j,\alpha_j , 
{\vec p}_j, \Gamma_j, {\vec n}_j\rangle =0$ for $i\neq j$. 
It is then straightforward
to see that $\langle v|v\rangle$ may be evaluated using equations (\ref{ip1}) and (\ref{ognlsum}) to yield the manifestly
positive definite expression:
\begin{equation}
\langle v|v\rangle = \sum_{i=1}^m \int ds_i |D_i(s_i)|^2 .
\end{equation}
The same argument goes through even if some or all of the ${\alpha}_i$ correspond to the trivial graph.

The inner product can be used to complete $V_{new}$ to the Hilbert space ${\cal H}_{new}$.

\subsection{Representation of operators on the Hilbert space.}
In this section we demonstrate the validity of the assumption (\ref{assumeo}) in the context of the 
new representation for abelian gauge fields. We define the action of the operators ${\hat h}_{\alpha, {\vec p}}$,
${\hat E}_{{\cal S}, f}$ and ${\hat U}_d$ in accordance with section 2. The holonomy operator acts as follows.
\begin{equation}
{\hat h}_{\alpha, {\vec p}}|C,\beta , {\vec q}, \Gamma , {\vec n}\rangle
= |C, \beta \cup \alpha, {\vec q}\cup{\vec p},\Gamma , {\vec n}\rangle ,
\label{hhatnew}
\end{equation}
\begin{equation}
{\hat h}_{\alpha, {\vec p}}|C, \Gamma , {\vec n}\rangle
= |C, \alpha, {\vec p},\Gamma , {\vec n}\rangle .
\label{hhatnew1}
\end{equation}
Clearly this action satisfies the assumption (\ref{assumeo}). The flux operator acts as follows.
\begin{equation}
{\hat E}_{{\cal S},f}
|C,\alpha , {\vec p},\Gamma, {\vec n}\rangle
=\hbar |CC^{{\cal S},f}_{\alpha, {\vec p}, \Gamma , {\vec n}},\alpha , {\vec p},\Gamma, {\vec n}\rangle ,
\label{ehatnew}
\end{equation}
\begin{equation}
{\hat E}_{{\cal S},f}
|C,\Gamma, {\vec n}\rangle
=\hbar |CC^{{\cal S},f}_{\Gamma , {\vec n}},\Gamma, {\vec n}\rangle 
\label{ehatnew1}
\end{equation}
where
\begin{equation}
C^{{\cal S},f}_{\alpha, {\vec p}, \Gamma , {\vec n}}(s) =  C^{{\cal S},f}_{\Gamma , {\vec n}}(s) \;+\;
( \sum_{i_{\alpha}} f(x_{i_{\alpha}}) \sum_{I_{i_{\alpha}}}p_{I_{i_{\alpha}}}
\kappa_{I_{i_{\alpha}}}) ,
\end{equation}
\begin{equation}
C^{{\cal S},f}_{\Gamma , {\vec n}}(s)= 
(\sum_{i_{s}} f(x_{i_{s}}) \sum_{I_{i_{s}}}n_{I_{i_{s}}}
\kappa_{I_{i_{s}}}).
\end{equation}
The notation is similar to that in equation (\ref{pb1}) and the subscripts $\alpha, s$ are used to
designate indices appropriate to the graphs $\alpha, \gamma (s)$ and their intersections with 
$\cal S$.  Thus, for example, $i_s$ ranges over the (transverse) intersections of $\gamma (s)$ with ${\cal S}$
and $I_{i_{s}}$ over the (transverse) edges of $\gamma (s)$ ending/begining at the interesction point $x_{i_{s}}$.
If $C^{{\cal S},f}_{\alpha, {\vec p}, \Gamma , {\vec n}},\;C^{{\cal S},f}_{\Gamma , {\vec n}}$ are
 piecewise smooth, then the 
right hand sides of the above equations are in ${\cal B}$. Hence we need to show that 
$C^{{\cal S},f}_{\Gamma , {\vec n}}$
is a piecewise smooth function.

Note that the surface ${\cal S}$ can intersect the graph $\gamma (s)$ at most a finite number of times.
Denote the $I$th edge of $\gamma (s) $ by $e_I(s)$ and denote the surface traced out by $e_I(s)$ as $s$ varies, by 
${\cal S}_I$. 
Lemma 1 of the Appendix implies that ${\cal S}_I$ is an analytic surface with piecewise
analytic boundary. Note that both ${\cal S}$ and ${\cal S}_I$ are semianalytic surfaces (see Reference \cite{lost}
and the references therein for a definition of semianalyticity). It follows from Reference \cite{lost}
that ${\cal S}\cap {\cal S}_I$ is the union of a finite number of (a) isolated points, (b) piecewise analytic 
curves, and (c) 2 dimensional semianalytic surfaces. Isolated points are of measure zero in $s$ space and can
therefore be ignored. The contributions from (c) to $C^{{\cal S},f}_{\Gamma , {\vec n}}$
 vanish since the associated edges are 
tangential to ${\cal S}$. Consider an analytic segement $\tau$ of (b) with analytic parameterization
$u$ so that $\tau$ traces out the curve $(t_I(u), s(u))$ on ${\cal S_I}$ where $t_I$ is the parameter along
the edge $e_I$. By Lemma 1, $s(u)$ is an analytic function. Hence either \\
\noindent (1) ${\frac{ds}{du}}=0$ at a finite number of points, or,\\
\noindent (2) $\tau$ is along some edge $e_I(s)$. \\
Case (2) is again one of tangential intersection and does not contribute to 
$C^{{\cal S},f}_{\Gamma , {\vec n}}$.
Since the only nontrivial contributions are from Case (1) and since the number of edges of $\gamma (s)$ is
finite (and independent of $s$), it is straightforward to see that 
$C^{{\cal S},f}_{\Gamma , {\vec n}}$
is a bounded, piecewise constant
function of $s$. 

Finally the operator ${\hat U}_d$ acts as follows:
\begin{equation}
{\hat U}_d |C,\beta , {\vec q}, \Gamma , {\vec n}\rangle
=|C,\beta_d , {\vec q}, \Gamma_d , {\vec n}\rangle ,
\label{diffhatnew}
\end{equation}
\begin{equation}
{\hat U}_d |C,\Gamma , {\vec n}\rangle
=|C,\Gamma_d , {\vec n}\rangle .
\label{diffhatnew1}
\end{equation}
Here $\Gamma_d$ is the equivalence class of the 1 parameter set of graphs ${\gamma}_d (s)$, where 
$\gamma_d (s)$ is obtained by the action of $d$ on $\gamma (s)$ and the labelled graph $\beta_d, {\vec q}$ 
is the image of the labelled graph $\beta , {\vec q}$ by $d$.
Note that if $\gamma (s)$ is obtained by 
the action of the diffeomorphism $\phi_{\xi} (s)$ on some graph $\gamma$ for some analytic vector field $\xi$, 
it follows that 
$\gamma_d (s)$ is obtained by the action of $\phi_{\xi_d} (s)$ on $\gamma_d$ where $\xi_d$ is the analytic vector field
obtained by the action of $d$ on $\xi$ and $\gamma_d$ is the image of $\gamma$ by $d$.
Thus, the right hand side of equation (\ref{diffhatnew}) is also in ${\cal B}$ and the assumption (\ref{assumeo})
is valid.

The reader may check explicitly that both the adjointness
\footnote{
As mentioned in section 3.2,  $V_{new}$ is completed to the Hilbert space ${\cal H}_{new}$. Since 
${\hat h}_{\alpha , {\vec p}}, {\hat U}_d$ are bounded operators, they admit unitary extensions
to all of ${\cal H}_{new}$. However ${\hat E}_{{\cal S}, f}$
 is unbounded and hence only densely defined
with dense domain $V_{new}$. While 
it is straightforward to check that ${\hat E}_{{\cal S}, f}$  is symmetric on this domain, we do not address issues of 
self adjointness here other than to remark that it should be much easier to handle the bounded operators
$\exp (i {\hat E}_{{\cal S}, f})$.}
and algebraic relations on these operators are 
represented through equations (\ref{ip1}),(\ref{hhatnew}),(\ref{hhatnew1}), 
(\ref{ehatnew}), (\ref{ehatnew1}) 
and (\ref{diffhatnew}),(\ref{diffhatnew1}).

\subsection{Open Issues and Remarks.}
\subsubsection{Infinitesmal and finite diffeomorphisms, ${\cal U}$ and ${\cal UD}$.}

Let the analytic vector field $\xi$ generate the one parameter family of diffeomorphisms $\phi_{\xi}(s)$.
As shown below, the operators ${\hat U}_{\phi_{\xi}(s)}$ do not have the recquisite continuity properties in
$s$ to define their generator as an operator on the dense domain $V_{new}$. Despite the inability to define 
such an operator on $V_{new}$, a more limited notion of infinitesmal diffeomorphisms does exist. Consider
the state $|C, \Gamma, {\vec n}\rangle$ and let $\gamma (s)$ be obtained by dragging $\gamma$ along the orbits 
of the vector field $\xi$. As in section 3.2, let 
$|C, \{\gamma (s),{\vec n}, s\in {\bf S}\}\rangle$ be in the equivalence class of $|C, \Gamma, {\vec n}\rangle$.
Further, let $C(s)$ be smooth and let it be supported in ${\bf S} = [s_1,s_2]$ for some $s_1,s_2 \in R$. 
From equation (\ref{diffhatnew1}) it follows that for small enough $\delta$,
\begin{eqnarray}
{\hat U}_{\phi_{\xi}(s=\delta)}|C, \{\gamma (s),{\vec n}, s\in [s_1,s_2]\}\rangle
=|C, \{\gamma (s+\delta),{\vec n}, s\in [s_1 ,s_2 ]  \}\rangle &\nonumber \\
=|C_{\Delta}, \{\gamma (s),{\vec n}, s\in [s_1+\delta ,s_2+\delta ]\}\rangle ,&
\end{eqnarray}
where $C_{\Delta}(s):= C(s-\delta)$. Applying equation (\ref{identify2a}), we have that 
\begin{eqnarray}
({\hat U}_{\phi_{\xi}(\delta)}- 1)|C, \{\gamma (s),{\vec n}, s\in [s_1,s_2]\}\rangle
&= &|C_{\Delta} -C, \{\gamma (s),{\vec n}, s\in [s_1 +\delta ,s_2]\}\rangle \nonumber\\
+|C_{\Delta}, \{\gamma (s),{\vec n}, s\in [s_2,s_2+\delta]\}\rangle 
&-& |C, \{\gamma (s),{\vec n}, s\in [s_1,s_1+\delta]\}\rangle \nonumber\\
= \delta |C^{\prime}, \{\gamma (s),{\vec n}, s\in [s_1,s_2]\}\rangle 
&+& \delta^2 |\psi\rangle ,
\label{inftsml}
\end{eqnarray}
where $|\psi\rangle$ has norm of $O(1)$ as $\delta \rightarrow 0$ and where $C^{\prime}$ is a half density 
which, in the parametrization $s$, evaluates to 
$C^{\prime}(s) := \frac{\partial C(s)}{\partial s}$. Equation (\ref{inftsml}) implies that 
\begin{equation}
\lim_{\delta\rightarrow 0}\frac{({\hat U}_{\phi_{\xi}(\delta)}- 1)}{\delta}
|C, \Gamma, {\vec n}\rangle
=|C^{\prime}, \Gamma , {\vec n}\rangle .
\label{limited}
\end{equation}
Equation (\ref{limited}) is the limited notion of infinitesmal diffeomorphisms which exists in the new representation.
Note that for $\alpha$ such that $\alpha (s) := \phi_{\xi}(s) \alpha \neq \alpha$ for all $s$ in some neighbourhood
of the origin, the limit
\begin{equation}
\lim_{\delta\rightarrow 0}\frac{({\hat U}_{\phi_{\xi}(\delta)}- 1)}{\delta}
|C,\alpha, {\vec p}, \Gamma , {\vec n}\rangle,
\end{equation}
does not exist due to the fact that 
$\langle C,\alpha, {\vec p}, \Gamma , {\vec n} |{\hat U}_{\phi_{\xi}(\delta)}
|C,\alpha, {\vec p}, \Gamma , {\vec n}\rangle = 0 $ for all small enough $\delta$.
This shows that $\lim_{\delta\rightarrow 0}\frac{({\hat U}_{\phi_{\xi}(\delta)}- 1)}{\delta}$ does not
exist on all of $V_{new}$.

We do not know if an operator for infinitesmal diffeomorphisms can  be defined  on some other dense domain in 
${\cal H}_{new}$, but we consider it unlikely. We also feel that such an operator cannot be defined
as a limit of operators in the holonomy flux algebra but this needs to be shown. Likewise,
while we feel that the  finite diffeomorphism operator ${\hat U}_d$ cannot be defined as a limit of operators
in the holonomy- flux algebra, this too remains an open question. In the unlikely event that such an operator
can be defined in this way, our intuition is that there must be at least one such definition which corresponds
to the action of ${\hat U}_d$ as defined in equations (\ref{diffhatnew})- (\ref{diffhatnew1}). If 
our expectations our correct, the extension of ${\cal U}$ to ${\cal UD}$ and the subsequent analysis of this
work is, we believe, fully justified.

\subsubsection{Cyclicity, GNS states and the flux operators.}

The new representation is not cyclic but cyclic subspaces can be defined via the standard Gelfand- Naimark- Segal
 (GNS) construction (see for example \cite{GNS}). Any state  $\Psi \in {\cal H}_{new}$ defines, via its
expectation values, a positive linear functional (PLF), on ${\cal UD}$ (or any of its subalgebras).
This PLF defines a cyclic representation (via the GNS construction) in which $\Psi$ is a cyclic state.
Let $\Psi = |C, \Gamma, {\vec n}\rangle$ be a normalised state (so that $\int ds |C(s)|^2 =1$) and 
consider the associated PLF evaluated on the commutative algebra of holonomies, $HA \subset {\cal UD}$.
Clearly
\begin{eqnarray}
\langle C, \Gamma, {\vec n}| {\hat h}_{\alpha , {\vec p}}|C, \Gamma, {\vec n}\rangle &=& 0 \;\forall\alpha \neq \circ
\nonumber\\
&=& 1 \;\;\;{\rm for}\; \alpha = \circ .
\label{plfha}
\end{eqnarray}
This is exactly the standard PLF on $HA$ used to construct the flux net representation.
The difference with the standard PLF arises when one considers the algebra $\cal U$ of holonomies and fluxes.
By using the commutators (\ref{pb1})- (\ref{pb2}), any element of $\cal U$ can be expressed as a linear combination
of terms, each of the form ${\hat h}_{\alpha , {\vec p}}\prod_{i=1}^m  {\hat E}_{{\cal S}_i, f_i}$.
From equations (\ref{ehatnew}), (\ref{ehatnew1}) and (\ref{ip2}), we have that 
\begin{eqnarray}
\langle C, 
\Gamma, {\vec n}| {\hat h}_{\alpha , {\vec p}}\prod_{i=1}^m  {\hat E}_{{\cal S}_i, f_i}|C, \Gamma, {\vec n}\rangle
&=& 0 \\
\langle C, 
\Gamma, {\vec n}| \prod_{i=1}^m  {\hat E}_{{\cal S}_i, f_i}|C, \Gamma, {\vec n}\rangle
&=& \int ds |C(s)|^2 \prod_{i=1}^m C^{{\cal S}_i,f_i}_{\Gamma , {\vec n}} .
\label{plfe}
\end{eqnarray}
In the standard flux network representation, any state in ${\cal H}$ is a linear combination of 
at most a {\em countable} infinity of flux network states. In contrast, equation (\ref{plfe}) refers to
the structure provided by the {\em uncountably} many graphs ${\gamma (s)}$.

Finally, the PLF evaluated on the remaining part of ${\cal UD}$ can be obtained by using equations
(\ref{ud1})- (\ref{ud4}) in conjunction with its evaluation on elements of the form 
${\hat h}_{\alpha , {\vec p}}(\prod_{i=1}^m  {\hat E}_{{\cal S}_i, f_i}){\hat U}_d$. It is straightforward to
evaluate this and we do not display the result here, other than to comment that the PLF vanishes
unless $\alpha = \circ$.

\subsubsection{Reducibility.}
Denote the cyclic representation of ${\cal UD}$ considered above by $({\cal UD}, {\cal H}_{C, \Gamma , {\vec n}})$.
Here ${\cal H}_{C, \Gamma , {\vec n}}$ is the GNS Hilbert space associated with the PLF of section 3.4.2.
Irreducibility of $({\cal UD}, {\cal H}_{C, \Gamma , {\vec n}})$ is an involved issue and we do not address it in 
this work. However, it is straightforward to see that representation,
$({\cal U}, {\cal H}^{\cal U}_{C, \Gamma , {\vec n}})$, obtained 
if we restrict 
the PLF of section 3.4.2 to the holonomy- flux algebra
 is, most likely, infinitely reducible. From equation (\ref{hhatnew1}), the operator
${\hat h}_{\alpha , {\vec p}}$ does not alter the half density label $C$ in $|C, \Gamma, {\vec n}\rangle$.
However (see equation (\ref{ehatnew1})) the flux operator ${\hat E}_{{\cal S}, f}$ 
does alter $C$ to $C_1:= CC^{{\cal S}, f}_{\gamma , {\vec n}}$.
Clearly, by  choosing ${{\cal S}, f}$ appropriately $C_1$ can have smaller support than $C$ but can never have larger
support than $C$. Thus, the cyclic subspace generated by using the PLF defined by 
$|C_1, \Gamma, {\vec n}\rangle$ is invariant under the action of elements of  $\cal U$. Repeated action by 
appropriately chosen flux operators yield further cyclic subspaces which are invariant with respect to
$\cal U$ so that the GNS representation based on the PLF defined by $|C, \Gamma, {\vec n}\rangle$ and 
restricted to $\cal U$ is infinitely
reducible with respect to $\cal U$. Note, however, that this argument pertains only to the cyclic subspace
generated by the action of  $\cal U$ on $|C, \Gamma, {\vec n}\rangle$. This subspace is dense in (and not equal to)  
the GNS Hilbert space
${\cal H}^{\cal U}_{C, \Gamma , {\vec n}}$. Hence, the argument as it stands, 
is not, strictly speaking, complete in that it 
does not adequately address issues of adjointness of the densely defined, unbounded flux operators.
Thus, the notion of irreducibility in the context of $({\cal U}, {\cal H}^{\cal U}_{C, \Gamma , {\vec n}})$ is 
complicated by the fact that the flux operators are unbounded and hence only densely defined. However, it may be
possible to show infinite reducibility in the context of ${\cal H}^{\cal U}_{C, \Gamma , {\vec n}}$ if 
(similar to the idea in Reference \cite{irreducibleLQG}), ${\cal U}$ is replaced by the algebra generated
by holonomies and exponentials of electric fluxes (i.e. by the bounded, unitary operators
$e^{i {\hat E}_{{\cal S}, f}}$).

This would be an undesirable feature if we has access only to ${\cal U}$ 
(or its Weyl algebra- like replacement alluded to above). However the operators ${\hat U}_d$ map states 
in ${\cal H}^{\cal U}_{C, \Gamma , {\vec n}}$ out of  ${\cal H}^{\cal U}_{C, \Gamma , {\vec n}}$ 
and hence are not superselected
with respect to ${\cal H}^{\cal U}_{C, \Gamma , {\vec n}}$. Moreover, it is conceivable that some of these operators
(at least for some choices of $\Gamma$) may be used to ``stretch'' the support of $C$ by ``stretching''
the 1 parameter family of graphs ${\gamma (s)}$.  It would be interesting to see if this is indeed possible,
as it impinges on the issue of irreducibility of $({\cal UD}, {\cal H}_{C, \Gamma , {\vec n}})$.

\subsubsection{Inequivalence with the standard flux net representation.}

As noted in section 3.4.2, the flux operators have a very different action (based on a uncountable family of 
flux net labels) than in the standard representation (where it can depend at most on a countable infinity 
of flux net labels) and this difference shows their inequivalence. Yet another reason to believe that the 
representations are inequivalent can be traced to the nature of the eigen functions of the flux operators.
In the standard representation, every flux net state is an eigen state of the flux operators and every such state
is normalizable. In the new representation, equations (\ref{ehatnew})- (\ref{ehatnew1}) suggest that 
eigen states may be obtained by replacing $C(s)$ in those equations by  the Dirac delta function 
$\delta (s_0, s)$, where $\delta (s_0, s)$ is a half density in each of its arguments. Clearly, such a replacement
yields a non- normalizable state (also see Footnote \ref{dirac} in this regard). A third way to demonstrate 
inequivalence would be to show that no diffeomorphism invariant state exists in ${\cal H}_{new}$. Direct inspection
shows that there is no such state in $V_{new}$. Since any state 
$|C,\alpha , {\vec p},\Gamma, {\vec n}\rangle$ is associated with a graph and a finite number of analytic 
surfaces traced out by the edges of $\gamma (s)$ (see Lemma 1 in the Appendix), it follows that any state 
in ${\cal H}_{new}$ is associated with, at most, a countable infinity of graphs and surfaces. Since the number of
(finite) analytic diffeomorphisms is uncountably large, it seems unlikely that a diffeomorphism invariant state
exists in ${\cal H}_{new}$. It would be of interest to convert these arguments into rigorous proofs.

As noted in section 3.4.2 the PLF based on  $|C,\Gamma, {\vec n}\rangle$ restricted to the holonomy algebra $HA$ is 
identical to the standard PLF appropriate to the flux net representation. Hence there seems to be enough structure
to define the space ${\overline {\cal A}}$ of generalised connections \cite{aajurekprojlimit}. The flux operators
in the standard representation are related to derivations on this space; it would be of interest to see 
if the flux operators in the new representation have any interpretation in terms of structures on
${\overline {\cal A}}$.  This issue has a bearing on the discussion centered around $\bf (1)$ in section 1.

\section{Concluding Remarks.}

In this work we have constructed a new representation for a diffeomorphism invariant theory of abelian
connections. In the new representation, finite diffeomorphisms act unitarily. It is in this sense that 
the representation is ``background independent''. From the point of view of quantum states as positive linear functionals,
the key difference between the standard flux net representation
and the one constructed here is in the evaluation of the PLF  on  electric flux operators (\ref{plfe}).
There are a number of open questions regarding the new representation and we urge the reader to
peruse section 3.4 wherein they are described in detail. We emphasise once again that though the 
cyclic sector (see section 3.4.2) of the 
new representation 
is,  most likely, infinitely reducible 
when defined with respect to 
the holonomy- flux algebra ${\cal U}$ 
this is probably not the case with respect 
to the holonomy- flux- diffeomorphism algebra
${\cal UD}$.
\footnote{As indicated in section 3.4.2 a precise definition of irreducibility is only available for
algebras of bounded operators and hence, strictly speaking, the flux operators in ${\cal U}, {\cal UD}$
should be replaced by suitable bounded functions thereof (see section 3.4.2 for further discussion of this).}
 Moreover, if our intuition is correct, the operators ${\hat U}_d$ corresponding to finite diffeomorphisms
either cannot be defined as limits of operators in ${\cal U}$ or can be defined in terms of such limits in such a way
as to agree with their definition in the new representation. If this is true, then there is no reason not to 
take the algebra $\cal UD$ as a starting point for quantization rather than the algebra ${\cal U}$.

In this work we have restricted attention to the analytic category. However we expect our considerations to be 
robust enough to generalise easily to the semianalytic category as well.


We leave a generalization of our constructions to the case of gauge group $SU(2)$ for future work. 
Note that there may be implementations of the ideas of section 2 which are different from the 
sort of constructions in section 3. For example, one could attempt to  define a 1 parameter family of states 
in the non- abelian case by first considering a (non gauge invariant) spinnet based on a loop 
with a single analytic edge, a single vertex with an intertwiner and a vector in an appropriate representation
of the gauge group \cite{aajurekarea}, and then generating the 1 parameter family by moving the vertex and its
interwtwiner and vector labels along the loop - the parameter in this case is the position of the vertex.
It would be of interest to see if this (or other ideas) can lead to interesting new representations.
Finally, we hope that the work here, in particular the form of the GNS functional (\ref{plfha}), (\ref{plfe}),
may motivate the constructions of alternate representations for canonical quantum gravity.

\vspace{3mm}

\noindent{\bf Acknowledgments}:
This work could not have been completed without Hanno Sahlmann's help. I gratefully acknowledge crucial
discussions with him and thank him for his constant encouragement.
I also gratefully acknowledge an invitation from T. R. Seshadri to visit the Physics Department of
Delhi University where part of this work was completed.

\section*{Appendix}

\noindent {\bf Lemma 1} Let $\Sigma$ be a 
real analytic, 3 dimensional, compact manifold (without boundary). Let $e$ be a 
non- self intersecting analytic curve in $\Sigma$ i.e.
$e:T\rightarrow \Sigma $ is an injective analytic map from an open interval
$T$ of the real line $R$ into $\Sigma$. Let $\xi$ be an analytic vector field on $\Sigma$
and $U \subset \Sigma$ an open set  such that $\xi$ is non- vanishing in $U$ and
$e(t) \subset U \; \forall \;t \in T$. Let $\phi_{\xi} (s),\; s\in R$, denote the one 
parameter family of diffeomorphisms of $\Sigma$ generated by $\xi$ with 
$\phi_{\xi} (0)$ being the identity map. Let 
there exist an open neighbourhood,$S$, of the origin in $R$ 
such that $e(s,t) := \phi_{\xi} (s)e(t) \subset U$ $\forall \; s\in S, \; t\in T$. 
Further, let the image, $\phi_{\xi} (s) e$ of the edge $e$ under the diffeomorphism
$\phi_{\xi} (s)$ be transverse to $\xi$ $\forall s \in S$.

Then the set
\begin{equation}
S_{\xi, S, T}=\{e(t,s)\; \forall  s\in S, \; t\in T\}
\label{defs}
\end{equation}
is an analytic 2- surface in $\Sigma$ with analytic chart $(s,t)$.

\vspace{3mm}

\noindent {\bf Proof}: Standard results for the smooth category imply that 
$S_{\xi, S, T}$ is a smooth 2- surface with chart $(s,t)$. Clearly, given
$(t_0,s_0) \in T\times S$, there exist open neighbourhoods $V_S\subset S$, $V_T\subset T$
and $V\subset \Sigma$ such that:\\
\noindent (a) $(t_0,s_0)\in V_T\times V_S$,\\
\noindent (b) $V$ is covered by a single analytic chart,\\
\noindent (c) $\{ e(t,s) \forall\; (t,s) \in V_T \times V_S\} \subset V$

Since 
$\phi_{\xi} (s)$ is a 1 parameter family of {\em analytic} diffeomorphisms
\footnote{While it seems to be standard folklore that analytic vector fields
generate analytic diffeomorphisms, we are unable to locate  this
result in the literature. However, we have verified that (in the context of compact manifolds
without boundary, where results for the smooth category hold) an application of the Cauchy- Kowalewsky
theorem \cite{CKthm} proves the result.}
and $e(t)$ is an analytic curve, it follows that $e^{\mu}(t,s_0),\; t\in V_T, \;\mu=1,2,3$ are 
analytic functions of $t$ (here $e^{\mu}(t,s)$ are the coordinates of $(e(t,s)$ in the analytic 
chart on $V$).

The surface $S_{\xi, S, T}$ is defined by the ``evolution'' equations 
\begin{equation}
\frac{\partial e^{\mu}}{\partial s} = \xi^{\mu}(e(t,s))
\label{I}
\end{equation}
and the initial conditions 
\begin{equation}
e^{\mu}(t,0)= e^{\mu}(t)
\label{II}
\end{equation}
Our strategy is to show that equations (\ref{I}) with initial data (\ref{II}) admit 
unique analytic solutions. This follows directly from an application of the 
Cauchy- Kowalewsky theorem \cite{CKthm}. In order to apply the theorem in the form
specified in Reference \cite{CKthm}, we proceed as follows.

We set $e^{\mu} (t_0, s_0)= 0$.
\footnote{That this entails no loss of generality follows from the fact that translation 
by a constant is an analytic transformation.}
Define the new coordinates $t^{\prime}:= t- t_0$ on $U_T$ and $s^{\prime}= s- s_0$ on $U_S$.
Define 
\begin{equation}
u_i= e^{\prime i}(t^{\prime}, s^{\prime}):=e^i(t, s), \;\;\; i=1,2,3.
\end{equation}
\begin{equation}
u_4 := t^{\prime}, \; x^1:= t^{\prime}.
\end{equation}
Define the functions $F_{i,j,k}(u_1,u_2,u_3,u_4), \; i,j=1,..,4, \; k=1$ as follows
\begin{eqnarray}
F_{i,j,1} &=& 0 \; {\rm for}\; i,j=1,2,3. \\
F_{i,4,1} &=& \xi^i (u_1,u_2,u_3)\; i=1,2,3. \\
F_{4,i,1}&=& 0 \; {\rm for} \; i=1,2,3,4.
\end{eqnarray}

It is straightforward to check that 
the content of the evolution equations (\ref{I}) with initial data (\ref{II}) can now be 
re-expressed in the evolution equations
\begin{equation}
\frac{\partial u_i}{\partial s^{\prime}}
= \sum_{j=1}^4 F_{i,j,k} \frac{\partial u_j}{\partial x^k}\; \;\;\;\;\;\;\;\; i=1,2,3,4 \;\;\;\;\; k=1
\label{III}
\end{equation}
with initial data 
\begin{equation}
u_i (t^{\prime}, 0) = e^{\prime i} (t^{\prime},0)= e^i (t,s_0) \;\; i=1,2,3 
\label{IV} 
\end{equation}
\begin{equation}
u_4(t^{\prime}, 0)= t^{\prime} .
\label{V}
\end{equation}
The analyticity of $F_{i,j,k}$ in the neighbourhood of the origin in $R^4$ follows from the 
analyticity of $\xi^{\mu}$ ; the analyticity of the initial data  (\ref{IV}) follows from the 
analyticity of the curve $e(t,s_0)$ (and the fact that the coordinates $t$ and $t^{\prime}$ as well as $s$ and $s^{\prime}$ 
are analytic functions of each other)
whereas the initial data (\ref{V}) is trivially analytic.
Finally, it can be checked that the initial data (\ref{IV})- (\ref{V}) vanish at the origin 
$(t^{\prime}, s^{\prime})= (0,0)$. 
Thus all the conditions of Theorem 2.4.1 of \cite{CKthm} are met and the equations
(\ref{III}) - (\ref{V}) are identical to  equations (2.43) and (2.44) of \cite{CKthm}.
It follows that there exist solutions  $u_i(t^{\prime}, s^{\prime}), i=1,..,4$ in a neighbourhood
of the origin $(t^{\prime}, s^{\prime})= (0,0)$ which are analytic functions of 
$(t^{\prime}, s^{\prime})$. Since $t= t^{\prime} +t_0,s= s^{\prime} +s_0$ are analytic functions
of $t,s$, it follows that there exists an open neighbourhood of $ (t=t_0,s= s_0)$  where 
$e^{\mu}(s,t),\; \mu=1,2,3$ are analytic functions of $(s,t)$. Since $(t_0,s_0)$ is an arbitrary 
point on $S_{\xi, S, T}$, this completes the proof of the Lemma.

\end{document}